\documentclass[12pt]{article}
\usepackage[dvips]{graphicx}

\makeatletter
\@addtoreset{equation}{section}

\renewcommand{\vec}[1]{\ensuremath{\mathchoice
                     {\mbox{\boldmath$\displaystyle#1$}}
                     {\mbox{\boldmath$\textstyle#1$}}
                     {\mbox{\boldmath$\scriptstyle#1$}}
                     {\mbox{\boldmath$\scriptscriptstyle#1$}}}}
\begin{document}
\begin{titlepage}
\setcounter{page}{0}
\begin{flushright}
{\baselineskip=12pt
Kagoshima HE-01-1\\
January, 2001

}
\end{flushright}

\vspace{1.5cm}
\begin{center}
{\baselineskip=30pt
{\Large\bf
Evolution of the quark-gluon and hadronic fluid 
with the compressible bag model
}

}

\vspace{1cm}
{\large
Shigenori {\sc Kagiyama}$^1$%
\footnote{e-mail: {\tt kagiyama@sci.kagoshima-u.ac.jp}}, 
Akira {\sc Minaka}$^2$%
\footnote{e-mail: {\tt minaka@edu.kagoshima-u.ac.jp}}
and 
Akihiro {\sc Nakamura}$^1$%
\footnote{e-mail: {\tt nakamura@sci.kagoshima-u.ac.jp}}

\bigskip
{$^1$ {\sl Department of Physics, Faculty of Science, Kagoshima University, 
Kagoshima 890-0065, Japan}}

{$^2$ {\sl Department of Physics, Faculty of Education, Kagoshima University, 
Kagoshima 890-0065, Japan}}

}
\end{center}

\vspace{2cm}
\begin{abstract}
\normalsize\noindent
The evolution of matter produced in nucleus collisions is 
investigated in the frame work of the simplified fluid equation, 
where the compressible bag model is used as the equation of state 
for hadrons.    It is shown that the quark-gluon phase is easily 
achieved in RHIC experiments, and that the typical structure of 
the phase transition, however, may be difficult to observe 
in simple analyses of $\langle p_t\rangle$ data.
\end{abstract}

\end{titlepage}

%%%%%%%%%%%%%%%%%%%%%%
\section{Introduction}
\label{intro}
There is long expectation to find the signal of quark-gluon plasma 
formation in the high energy experiments of nucleus-nucleus collision.  
Indeed, not a few of the experimental results suggest the existence 
of the quark phase \cite{proc}, but it seems that the incident energy is 
a little lower to observe the clean signals even at CERN SPS.  
The recent experiment at BNL RHIC may give us cleaner and richer 
signals.  

The purpose of the present paper is to give a manageable model 
for evolution of hadronic matter and quark-gluon plasma, produced 
in nucleus collisions, and to give some numerical predictions.  
There are two main ingredients in the model; 
the evolution equation of the matter and the equation of state 
for hadrons.

In the evolution equation, assumed is cylindrical symmetry with 
respect to the collision axis, and used are the averaged quantities 
over the transverse plane.  The averaged evolution equation is 
already reported in \cite{crude} \cite{fluid}, and its benefits 
are in that the parameterization of the unknown initial-conditions 
becomes very simple and in that the concerned physical quantities, 
such as size and transverse velocity of the matter, are directly 
obtained, in contrast to the straightforward calculation of the 
relativistic fluid equation \cite{3Dfluid}.

As for the equation of state for hadrons, the compressible bag 
model \cite{cmpb} \cite{crit} \cite{dibr} is used, which has nice 
features as explained in the following.  
The simplest point-like pion gas model is often 
used but is lack of degree of freedom, and then it cannot reproduce 
the observed multiplicity of hadrons as long as natural parameters 
are employed.  On the other hand, if the higher resonances are 
fully introduced in the point-like model, there appears the 
limiting temperature and the deconfinement transition to quark 
phase never takes place.  The compressible bag model gives a 
solution for this dilemma, which gives sufficient degree of 
freedom and sufficiently high temperature because the mass 
of the hadrons becomes heavier as they are compressed by 
surrounding pressure.  In this model the phase diagram with 
baryons is also calculable explicitly for whole region of 
parameters, and it coincides with the expected shape 
\cite{proc} \cite{dibr}.  

\begin{figure}
\begin{center}
\includegraphics[width=10truecm]{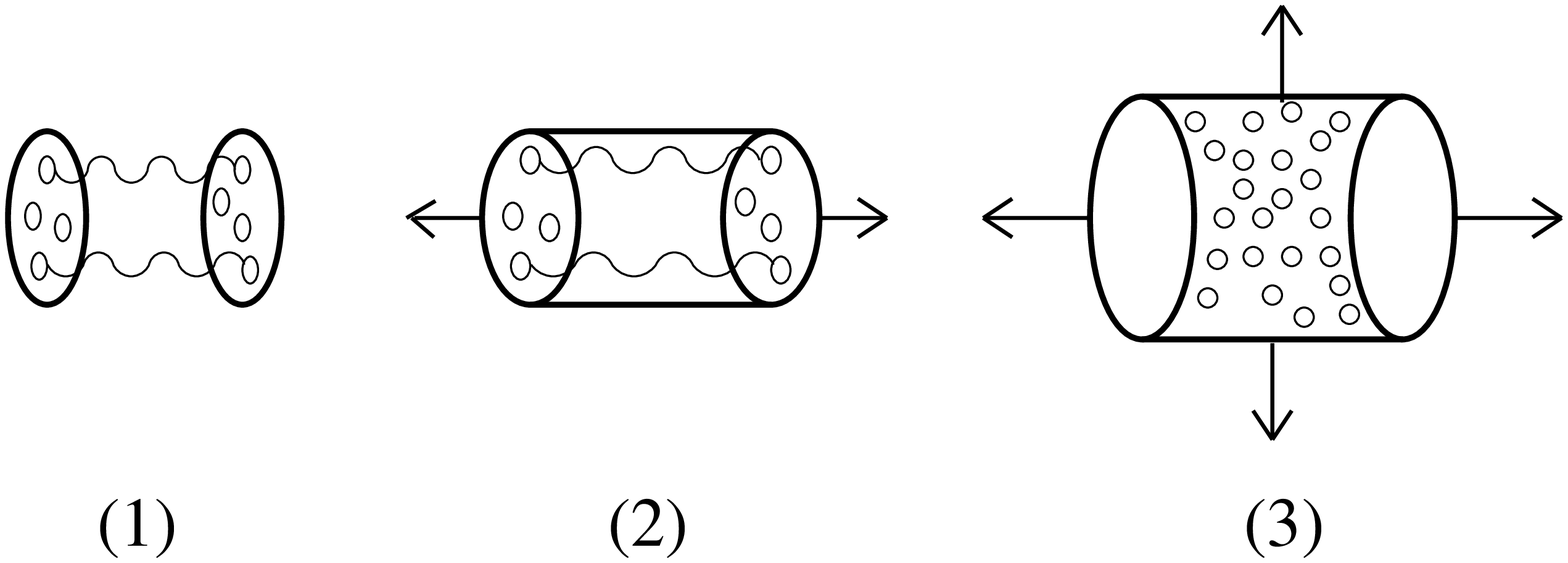}
\caption[]{(1) Step 1. Gluon exchange between nuclei.
(2) Step 2. Longitudinal expansion of the flux tube.
(3) Step 3. Cylindrical expansion of the hadronic or quark matter.}
\end{center}
\end{figure}
In Fig.~1, shown are the three steps in the nucleus collisions: 
At the first step, gluons are exchanged between the nucleons, 
and the both nucleus are shifted to higher color-charge state.  
The initial energy density is essentially given at this stage.  
In the second step, the color flux is expanded in the longitudinal 
direction, and at proper time $\tau_0\sim 1\,{\rm fm}$, the flux 
materializes into hadrons, quark-gluons or their mixed phase
depending on the energy density.  In the third step, the 
matter expands both in longitudinal and transverse directions,
being cooled down to the freeze-out temperature, and then breaks 
into individual hadrons.

Since the present model treats the third step, in order to 
show explicit results of the model, we must take some assumptions 
for the initial energy density of the produced fluid.  Therefore 
the details of the results depends on the assumption, but it is 
shown that the quark-gluon phase is achieved in RHIC experiments 
and that the signal of the phase transition, however, is difficult 
to observe in simple analyses of $\langle p_t\rangle$ data.  
The latter is mainly because of the richness of freedom in 
hadron phase: The richness pushes up the entropy density in hadron 
phase, and its difference between hadron and quark phases become 
much smaller than that calculated in simple pion gas model.  
Therefore the period of mixed phase becomes also smaller, and the 
typical structure of the first order phase transition is smeared out.

In the next section, the equation of state in the compressible 
bag model is briefly reviewed on the basis of the results of 
\cite{crit} \cite{dibr}.  In \S 3, evolution equations are 
presented for perfect fluid of hadron or quark-gluon or their 
mixture and it is revealed that the evolution equations have an 
expanding solution as well as a scaling solution.  And relations 
to observables are also given.  In \S 4, the results of numerical 
calculations are given and discussed on the basis of the expanding 
solution for $NN$-collisions and $AA$-collisions.  
Final section is devoted to conclusions and discussion.  

%%%%%%%%%%%%%%%%%
\section{Equation of states}
\label{eos}
In this section, we briefly summarize the results of \cite{crit} 
\cite{dibr}.  Let us suppose a mixed gas of $n$-kinds of meson 
enclosed in volume $V$ at finite temperature $T$.  
In the compressible bag model, the free energy function $\hat F$ 
of the gas is given by 
\begin{eqnarray}
\hat F &=& \sum_{i=1}^{n}F_f(N_i,V',T,M_i(v_i)), \label{eq:F}\\
V' &=& V-b\sum_{j=1}^{n}N_jv_j, \label{eq:Vex}
\end{eqnarray}
where $N_i$ is a number of $i$-th kind of meson and $v_i$ is its volume 
and $M_i$ its mass.  The constant $b$ is a volume exclusion efficiency 
parameter.   The function $F_f$ is a free energy function of free 
(point-like) boson gas.  As for mass function $M_i(v_i)$, we assume that 
of MIT bag model:  
\begin{eqnarray}
M_i(v_i) = A_iv_i^{-1/3}+Bv_i, \label{eq:MIT}
\end{eqnarray}
where $B$ is bag constant.  

Under the approximation that the average of inverse Lorentz factor
of $i$-th kind of meson equals to unity ($\langle\gamma_i^{-1}
\rangle\approx 1$) in the rest frame of the system, 
basic requirement of the compressible bag model 
that $\partial\hat F/\partial v_i=0$ and the requirement that the 
the chemical potential of meson should vanish, 
$\mu_i=\partial\hat F/\partial N_i=0$, determine the pressure $p$ 
of the system as a function of $T$ by the following equation: 
\begin{eqnarray}
p &=& -\sum_{i=1}^{n}g_iT\int{{d^3\vec{k}}\over{(2\pi)^3}}
\ln\left\{1-\exp[-(E_i-\mu'_i(p,m_i))/T]\right\}, \label{eq:pi}\\
&&E_i=\sqrt{\vec{k}^2+M_i(p,m_i)^2},\cr
M_i(p,m_i) &=& m_i\left(1+{{3bp}\over{4B}}\right)\left(1+{{bp}\over{B}}
\right)^{-3/4}, \label{eq:Mi}\\
\mu'_i(p,m_i) &=& -bv_ip=-{{bm_ip}\over{4B}}\left(1+{{bp}\over{B}}
\right)^{-3/4} \label{eq:vi},
\end{eqnarray}
where $g_i$ is a degeneracy factor of the $i$-th kind of meson and 
$m_i=4(A_i/3)^{3/4}B^{1/4}$ is its mass in the vacuum.  

Now we take the limit $n\to\infty$ and replace the sum with the integral
\begin{eqnarray}
\sum_{i=1}^{n}g_i \Longrightarrow \int dm\,\tau(m), \label{eq:limit}
\end{eqnarray}
where $\tau(m)$ is a level density function.
Then (\ref{eq:pi}) is written as
\begin{eqnarray}
p &=& - \int dm\,\tau(m)T\int{{d^3\vec{k}}\over{(2\pi)^3}}
\ln\left[1-e^{-(E-\mu')/T}\right], 
\quad E=\sqrt{\vec{k}^2+M^2}, \label{eq:p}\\
M &=& m\left(1+{{3bp}\over{4B}}\right)\left(1+{{bp}\over{B}}\right)^{-3/4},
      \label{eq:M}\\
\mu' &=& -{{bmp}\over{4B}}\left(1+{{bp}\over{B}}\right)^{-3/4}.
      \label{eq:mu}
\end{eqnarray}
Thus we can calculate $p$ as functions of $T$ from (\ref{eq:p}) 
when an appropriate level density function is given.
An entropy density $s$ is given by 
\begin{eqnarray}
s &=& {{dp}\over{dT}}={C\over D}, \label{eq:s}\\
C &=& {p\over T}+\int dm\,\tau(m)\int{{d^3\vec{k}}\over{(2\pi)^3}}
      {{(E-\mu')/T}\over{\exp[(E-\mu')/T]-1}}, \label{eq:C}\\
D &=& 1+\int dm\,\tau(m)\int{{d^3\vec{k}}\over{(2\pi)^3}}
     {1\over{\exp[(E-\mu')/T]-1}}
     {{d(E-\mu')}\over{dp}}, \label{eq:D}\\
{{d(E-\mu')}\over{dp}} &=& {{mb}\over{4B}}\left[1+\left(1+{{3M}\over E}
\right){{bp}\over{4B}}\right]\left(1+{{bp}\over{B}}\right)^{-7/4}.
\end{eqnarray}
An energy density $\varepsilon$ is calculated by the following 
thermodynamic relation 
\begin{eqnarray}
\varepsilon=Ts-p.  \label{eq:eps}
\end{eqnarray}
A number density $\rho$ for the hadron with vacuum mass $m$ is 
given by \cite{dibr}
\begin{eqnarray}
\rho &=& {{\rho'}\over{1+b\int dm\,\tau(m)\rho'v}}, \label{eq:rho}\\
v &=& {m\over{4B}}\left(1+{{bp}\over{B}}\right)^{-3/4}, \label{eq:v}\\
\rho' &=& \int{{d^3\vec{k}}\over{(2\pi)^3}}
             {1\over{\exp[(E-\mu')/T]-1}}. \label{eq:rhop}
\end{eqnarray}
Precisely speaking, the above $\rho$ is not the number density but 
one should multiply it by $\tau(m)$ and integrate over some mass 
range, say $[m-\Delta m, m+\Delta m]$, to get the number density 
of the hadron with vacuum mass around $m$.  
In practice, however, we use the above $\rho$ simply multiplied the 
degeneracy factor of the hadron, for example 3 for pion, instead of 
integrating, as the number density.  This prescription is justified 
when narrow width approximation is applicable, which we will 
assume in this paper where only pion density is considered in 
numerical calculations.  

As for the level density, we adopt the following form
\begin{eqnarray}
\tau(m) = 3\delta(m-m_\pi)+a\theta(m-2m_\pi)m^{-d}\exp(m/T_M), \label{eq:tau}
\end{eqnarray}
and whose parameters are
\begin{eqnarray}
a=25.46\,T_M^2, \quad d=3,\quad T_M=(90B/16\pi^2)^{1/4}, \label{eq:param1}
\end{eqnarray}
according to the Abelian bag model by Kapusta \cite{level}.

%%%%%%%%%%%%%%%%%
\section{A simple fluid model}
\label{fluid}
%%%%%%%%%%%%%%
\subsection{Evolution equations}
\label{evolution}
In this subsection, we present evolution equations of perfect fluid 
of hadron or quark-gluon or their mixture.  In the previous paper 
\cite{fluid}, hadrons had been assumed to be massless, but in the present 
paper, hadrons are assumed to obey the equation of states given in the 
previous section.  

As shown in \S3 of \cite{trans}, transverse flow with cylindrical 
symmetry obeys the following equations: 
\begin{eqnarray}
{{d\tilde s}\over{d\tau}} &=& -{{\tilde s}\over\tau}, \label{eq:St}\\
{{d\tilde\varepsilon}\over{d\tau}} 
&=& -{{\tilde\varepsilon+\tilde p}\over\tau}, \label{eq:Et}
\end{eqnarray}
where $\tau$ is defined by $(t^2 - z^2)^{1/2}$, and has the meaning
of the proper time only on the collision axis (z-axis).
$\tilde s$, $\tilde\varepsilon$ and $\tilde p$ are quantities 
integrated over the transverse plane; 
\begin{eqnarray}
\tilde s &=& \int d^2r\,s\gamma_t,\quad 
\tilde\varepsilon = \int d^2r\,[(\varepsilon+p)\gamma_t^2 - p],\quad
\tilde p = \int d^2r\,p,
\end{eqnarray}
where $\gamma_t$ is the transverse $\gamma$-factor of the fluid, 
and $\varepsilon$, $p$ and $s$ are energy density, pressure and 
entropy density respectively.  

Following \cite{crude}, let us introduce a radius $R$ of the 
cylindrical fluid and assume the following kinematical relation; 
\begin{eqnarray}
{{dR}\over{d\tau}} = v_t, \label{eq:dR}
\end{eqnarray}
where $v_t$ is the transverse velocity and is related to $\gamma_t$
by the relation $\gamma_t = (1-v_t^2)^{-1/2}$. And further let us 
take a crude approximation, 
\begin{eqnarray}
\tilde s &=& \pi R^2 s\gamma_t,\quad 
\tilde\varepsilon = \pi R^2 [(\varepsilon+p)\gamma_t^2 - p],\quad
\tilde p = \pi R^2 p, \label{eq:simple}
\end{eqnarray}
and hereafter regard $s$, $\varepsilon$ and $\gamma_t$ as the 
averages over the transverse plane, or as the values at the 
representative point of $r$.  All the quantities, then, only
depend on $\tau$.  

Integrating (\ref{eq:St}) and using (\ref{eq:simple}), 
we get the following equation 
\begin{eqnarray}
\tau\tilde s=\pi R^2 s \gamma_t\tau = {\rm const.}. \label{eq:tauS}
\end{eqnarray}
Differentiation of this equation with respect to $\tau$ yields 
\begin{eqnarray}
0&=&{{2\dot R}\over R}+{{\dot s}\over s}+{{\dot\gamma_t}\over{\gamma_t}}
    +{1\over\tau} \cr
\noalign{\smallskip}
 &=&{{2v_t}\over R}+{{\dot s}\over s}+{{v_t}\over{1-v_t^2}}\dot v_t
    +{1\over\tau}, \label{eq:dS}
\end{eqnarray}
where dot represents differentiation with respect to $\tau$.  

Before proceeding to derive an evolution equation, we should note 
the following relations obtained from the thermodynamic relations 
(\ref{eq:s}) and (\ref{eq:eps}), 
\begin{eqnarray}
{{\dot\varepsilon+\dot p}\over{\varepsilon+p}}
  ={{\dot T}\over T}+{{\dot s}\over s},\quad
{{\dot p}\over{\varepsilon+p}}={{\dot T}\over T},\quad 
{{\dot\varepsilon}\over{\varepsilon+p}}={{\dot s}\over s}. \label{eq:td}
\end{eqnarray}
Then using (\ref{eq:dR})$\sim$(\ref{eq:td}) and (\ref{eq:eps}), (\ref{eq:Et}) 
is rewritten as follows; 
\begin{eqnarray}
0&=&{{\dot{\tilde\varepsilon}}\over{\tilde\varepsilon+\tilde p}}
   +{1\over\tau} \cr
\noalign{\smallskip}
 &=&{{2\dot R}\over R}+{{\dot\varepsilon+\dot p}\over{\varepsilon+p}}
   +{{2\dot\gamma_t}\over{\gamma_t}}
   -{{2\dot Rp}\over{R(\varepsilon+p)\gamma_t^2}}
   -{{\dot p}\over{(\varepsilon+p)\gamma_t^2}}+{1\over\tau} \cr
\noalign{\smallskip}
 &=&{{2v_t}\over R}+{{\dot T}\over T}+{{\dot s}\over s}
   +{{2v_t}\over{1-v_t^2}}\dot v_t
   -{{2v_tp}\over{R(\varepsilon+p)\gamma_t^2}}
   -{{\dot T}\over{T\gamma_t^2}}+{1\over\tau} \cr
\noalign{\smallskip}
 &=&v_t\left\{ v_t{{\dot T}\over T}+{{\dot v_t}\over{1-v_t^2}}
   -{{2(1-v_t^2)p}\over{R(\varepsilon+p)}}\right\},
\end{eqnarray}
that is, 
\begin{eqnarray}
v_t=0,  \label{eq:sc}
\end{eqnarray}
or
\begin{eqnarray}
v_t{{\dot T}\over T}+{{\dot v_t}\over{1-v_t^2}}
-{{2(1-v_t^2)p}\over{R(\varepsilon+p)}}=0. \label{eq:ex}
\end{eqnarray}
The equation (\ref{eq:sc}) has a scaling solution $R=R_0$, 
$s=s_0\tau_0/\tau$, where $R_0$ and $s_0$ are initial values 
at $\tau=\tau_0$.  As for (\ref{eq:ex}), it together with (\ref{eq:dS}) 
is recasted as 
\begin{eqnarray}
\dot v_t &=& {{1-v_t^2}\over{1-\alpha v_t^2}}
  \left[ {{\alpha v_t}\over\tau}+{1\over R}
  \left\{ 2\alpha v_t^2+{{2p(1-v_t^2)}\over{\varepsilon+p}}
  \right\}\right], \label{eq:dvt}\\
{{\dot T}\over T} &=& -{{\alpha}\over{1-\alpha v_t^2}}
  \left[ {{2v_t}\over R}\left\{ 1+{{p(1-v_t^2)}\over{\varepsilon+p}}
  \right\}+{1\over\tau}\right], \label{eq:dT}\\
\alpha &\equiv& {s\over{Tds/dT}}.
\end{eqnarray}
The equations (\ref{eq:dR}), (\ref{eq:dvt}) and (\ref{eq:dT}) are 
a closed set of evolution equations of fluid, which is also 
applicable for the fluid in quark-gluon phase as well as hadron phase.  
These equations have an expanding solution if an initial value 
$v_t(\tau_0)$ is non-negative.  In the next section, we will numerically 
analyze on the basis of this expanding solution.  

For quark-gluon phase, equations of states are given by 
\begin{eqnarray}
p = K_Q T^4 - B,\quad s = 4 K_Q T^3,\quad
\varepsilon = 3 K_Q T^4 + B, 
\end{eqnarray}
with
\begin{eqnarray}
K_Q = \left({{7N_f}\over{60}}+{8\over{45}}\right)\pi^2, 
\end{eqnarray}
where $N_f$ is a number of quark flavor.  
The evolution equations (\ref{eq:dvt}) and (\ref{eq:dT}) 
for quark-gluon fluid \cite{fluid} read 
\begin{eqnarray}
\dot v_t &=& {{1-v_t^2}\over{3-v_t^2}}\left\{ {{v_t}\over{\tau}}
+{{3+v_t^2-3(1-v_t^2)B/K_Q T^4}\over{2R}} \right\}, \label{eq:dvtq}\\
{{\dot T}\over T} &=& -{1\over{3-v_t^2}}\left\{ 
{{v_t(5-v_t^2)}\over{2R}}+{1\over\tau} \right\}. \label{eq:dTq}
\end{eqnarray}

For mixed phase, equations of states are given by 
\begin{eqnarray}
p &=& p_c = p_Q(T_c) = p_H(T_c), \\
s &=& xs_Q(T_c) + (1-x)s_H(T_c), \\
\varepsilon &=& x\varepsilon_Q(T_c) + (1-x)\varepsilon_H(T_c),
\end{eqnarray}
where $T_c$ is critical temperature and the subscript $H$ ($Q$) 
denotes the quantity for hadron (quark-gluon), and $x$ is a 
fraction of quark-gluon.  Corresponding equations to (\ref{eq:dvt}) 
and (\ref{eq:dT}) are given by 
\begin{eqnarray}
\dot v_t &=& {{2p}\over{\varepsilon+p}}{{(1-v_t^2)^2}\over R},\label{eq:dvtm}\\
\dot x &=& {{\dot s}\over{s_Q(T_c)-s_H(T_c)}}
        = -{s\over{s_Q(T_c)-s_H(T_c)}} 
    \left[ {{2v_t}\over R}\left\{ 1+{{p(1-v_t^2)}\over{\varepsilon+p}}
    \right\}+{1\over\tau}\right]. \label{eq:dx}
\end{eqnarray}
Thus we have obtained evolution equations for all phases.  Those 
are (\ref{eq:dR}), (\ref{eq:dvt}) and (\ref{eq:dT}) for hadron phase, 
(\ref{eq:dR}), (\ref{eq:dvtq}) and (\ref{eq:dTq}) for quark-gluon phase 
and (\ref{eq:dR}), (\ref{eq:dvtm}) and (\ref{eq:dx}) for mixed phase.  
Throughout the evolution, $\tau\tilde s$ is conserved, as is seen in 
(\ref{eq:tauS}), so that it can be used as a check in solving 
the evolution equations numerically. 

%%%%%%%%%%%%%%
\subsection{Relation to observables}
\label{observable}
Now that we have the evolution equations, we can solve them and 
get $T$ ($x$ for mixed phase), $R$ and $\gamma_t$ as functions of 
$\tau$ if some initial values $T_0$, ($x_0$ for mixed phase) and 
$R_0$ are given and the initial condition $\gamma_t(\tau_0)=1$ is 
assumed.  
The final condition of the evolution is set by the freeze-out of 
hadrons from hadronic fluid.  If the freeze-out temperature $T_f$ is
given, we can get $\tau_f$ from the solution,
\begin{eqnarray}
T_f = T(\tau_f), \label{eq:Tf}
\end{eqnarray}
where the subscript $f$ means freeze-out.  
Then we get the values of the radius $R$ and $\gamma_t$ at the 
hadronization;
\begin{eqnarray}
R_f = R(\tau_f),\quad
\gamma_{tf} = \gamma_t(\tau_f). \label{eq:Rf} 
\end{eqnarray}
These quantities are related to experimental observables.  

\bigskip\noindent
(i) {\it Charged multiplicity}

The produced hadron density in rapidity space is given by
\begin{eqnarray}
{{dN}\over{dy}} = \tau_f\pi R_f^2\rho_f \gamma_{tf}, 
\end{eqnarray}
where $\rho_f$ is the number density at $\tau=\tau_f$.  
If we assume that all the produced hadrons are pions and put 
$N_{\rm ch}=(2/3)N$, we get the charged particle density,
\begin{eqnarray}
{{dN_{\rm ch}}\over{dy}} 
= {2\over 3}\tau_f\pi R_f^2\hat\rho_{\pi f}\gamma_{tf}, \label{eq:dNchdy}
\end{eqnarray}
where $\hat\rho_\pi$ denotes effective pion number density defined by 
\begin{eqnarray}
\hat\rho_\pi &=& \rho_\pi+N_M\rho_M, \label{eq:eff}\\
\rho_\pi &=& = {{\rho'_\pi}\over{1+\int dm\,\tau(m)\rho' v}}, \label{eq:rhopi}\\
\rho_M &=& {{\rho'_M}\over{1+\int dm\,\tau(m)\rho' v}}, \label{eq:rhoM}\\
\rho'_\pi &=& 3\int{{d^3\vec{k}}\over{(2\pi)^3}}
             {1\over{\exp[(E_\pi-\mu')/T]-1}}, \label{eq:rhoppi}\\
\rho'_M &=& \int_{2m_\pi}^{\infty}dm\,\tau(m)\int{{d^3\vec{k}}\over{(2\pi)^3}}
             {1\over{\exp[(E-\mu')/T]-1}}. \label{eq:rhopM}
\end{eqnarray}
In (\ref{eq:eff}), the first term corresponds to pole term and the 
second term corresponds to continuum part of $\tau(m)$ in 
(\ref{eq:tau}).  The quantity $N_M$ is an average multiplicity to 
pions from the continuum of meson resonances.  

\bigskip\noindent
(ii) {\it Transverse momentum and transverse energy}

The average transverse momentum of pion in the center of mass system 
is approximately given by 
\begin{eqnarray}
\langle p_t\rangle = \sqrt{\langle m_t^{*2}\rangle\gamma_t^2 - m_\pi^2},
\quad m_t^*=\sqrt{p_t^{*2}+m_\pi^2}, \label{eq:ptdef}
\end{eqnarray}
where $p_t^*$ ($m_t^*$) is a transverse momentum (mass) of pion in 
the rest frame of the fluid.  The average transverse mass of pion in 
the rest frame of the fluid is calculated as an average of the one for 
directly produced pion and the one for the contribution from pions 
as decay products of continuum of meson resonances; 
\begin{eqnarray}
\langle m_t^{*2}\rangle 
&=& {{\rho_{\pi f}\langle m_t^{*2}(\pi_d)\rangle
  +N_M\rho_{Mf}\langle m_t^{*2}(\pi_c)\rangle}
  \over{\rho_{\pi f}+N_M\rho_{Mf}}}. \label{eq:mtdef}
\end{eqnarray}
The quantity $\langle m_t^{*2}(\pi_c)\rangle$ is estimated on the basis of 
random walk approximation as 
\begin{eqnarray}
\langle m_t^{*2}(\pi_c)\rangle = \langle m_t^{*2}(M)\rangle/N_M. \label{eq:pic}
\end{eqnarray}
Then it follows from (\ref{eq:rhopi}), (\ref{eq:rhoM}), 
(\ref{eq:mtdef}) and (\ref{eq:pic}) 
\begin{eqnarray}
\langle m_t^{*2}\rangle
&=& {{\rho'_{\pi f}\langle m_t^{*2}(\pi_d)\rangle
  +\rho'_{Mf}\langle m_t^{*2}(M)\rangle}
  \over{\rho'_{\pi f}+N_M\rho'_{Mf}}}, \label{eq:mt}\\
\langle m_t^{*2}(\pi_d)\rangle
&=& { {\displaystyle{\int{{d^3\vec{k}}\over{(2\pi)^3}}
      {{k_t^2+M_\pi^2}\over{\exp[(E_\pi-\mu')/T]-1}}}} \over
   {\displaystyle{\int{{d^3\vec{k}}\over{(2\pi)^3}}
     {1\over{\exp[(E_\pi-\mu')/T]-1}}}} },\\
\langle m_t^{*2}(M)\rangle 
&=& { {\displaystyle{\int_{2m_\pi}^\infty dm\,\tau(m)
        \int{{d^3\vec{k}}\over{(2\pi)^3}}{{k_t^2+M^2}\over
        {\exp[(E_\pi-\mu')/T]-1}}}} \over
      {\displaystyle{\int_{2m_\pi}^\infty dm\,\tau(m)
        \int{{d^3\vec{k}}\over{(2\pi)^3}}
             {1\over{\exp[(E_\pi-\mu')/T]-1}}}} },
\end{eqnarray}
where $k_t$ denotes a transverse momentum variable.  

Transverse energy per unit rapidity interval is simply given by 
\begin{eqnarray}
{{dE_t}\over{dy}}
=\sqrt{\langle m_t^{*2}\rangle}\gamma_t{{dN}\over{dy}}.
\end{eqnarray}

\bigskip\noindent
(iii) {\it Rapidity interval and total multiplicity}

Since the evolution of fluid does not take place yet at $\tau=\tau_0$,
total energy of the fluid $E_{\rm tot}$ in the center of mass system 
is calculated as 
\begin{eqnarray}
E_{\rm tot} &=& \int_{-\eta_{\rm max}}^{\eta_{\rm max}}\tau_0\varepsilon_0
             \pi R_0^2\cosh\eta\,d\eta 
         = 2\tau_0\varepsilon_0\pi R_0^2\sinh\eta_{\rm max}.
\end{eqnarray}
Then the rapidity interval of the fluid $2\eta_{\rm max}$ is 
determined when $E_{\rm tot}$ and initial temperature $T_0$ 
($x_0$ for mixed-phase start) are given, that is, 
\begin{eqnarray}
2\eta_{\rm max} = 2\sinh^{-1}\left({{E_{\rm tot}}\over
                {2\tau_0\varepsilon_0\pi R_0^2}}\right). \label{eq:ymax}
\end{eqnarray}
Total multiplicity $\langle N_{\rm ch}\rangle$ is approximately given by 
\begin{eqnarray}
\langle N_{\rm ch}\rangle \approx 2\eta_{\rm max}{{dN_{\rm ch}}\over{dy}}. 
\end{eqnarray}

%%%%%%%%%%%%%%%%%
\section{Numerical calculations}
\label{calculations}
%%%%%%%%%%%
\subsection{$NN$-collision}
\label{NN}
In order to start numerical calculations, we have to fix bag constant 
$B$, volume exclusion parameter $b$, the number of quark flavors $N_f$, 
average multiplicity of pions from continuum $N_M$, initial proper time 
$\tau_0$, initial temperature $T_0$ ($x_0$ for mixed-phase start), initial 
radius $R_0$ and freeze-out temperature $T_f$.  First of all, we fix as 
\begin{eqnarray}
N_f = 2.
\end{eqnarray}
The initial radius $R_0$ for $NN$-collision is set to the experimental 
value of proton charge radius 
\begin{eqnarray}
R_0(NN)=0.82\,{\rm fm}. \label{eq:R0}
\end{eqnarray}
The parameter $b$ is determined by the relation 
\begin{eqnarray}
bv_N(p=0) = {{bm_N}\over{4B}} = {{4\pi}\over 3}(0.82\,{\rm fm})^3,\label{eq:Rc}
\end{eqnarray}
when the bag constant is given.  We adopt the following values 
\begin{eqnarray}
B^{1/4}=0.25\,{\rm GeV},\quad b=5.00. \label{eq:Bb}
\end{eqnarray}
Then critical values at the deconfinement transition are given by 
\begin{eqnarray}
T_c = 0.196\,{\rm GeV},\quad
\varepsilon_H(T_c)  = 0.874\,{\rm GeV/fm^3},\quad
\varepsilon_Q(T_c) = 2.85\,{\rm GeV/fm^3}. \label{eq:crt}
\end{eqnarray}

For the purpose of determining $T_f$ and $\tau_0$, we use the fit 
to experimental data of $NN$-collisions as inputs \cite{fit}; 
\begin{eqnarray}
T_f &=& 0.135\pm 0.003\,{\rm GeV}, \label{eq:Tffit}\\
\gamma_{tf} &=& 1+(0.010\pm 0.003)\log(E_{\rm cm}/E_1)
                \theta(E_{\rm cm}-E_1), \label{eq:gammafit}\\
&&E_1 = 12\pm 8\,{\rm GeV}, \cr
{{dN_{\rm ch}}\over{dy}} &=& (1.31\pm 0.02)
    +(0.20\pm 0.03)\log(E_{\rm cm}/E_2) \cr
  &&+(0.062\pm 0.007)\log^2(E_{\rm cm}/E_2), \label{eq:dNdyfit}\\
  &&E_2=12\,{\rm GeV},\nonumber
\end{eqnarray}
where $E_{\rm cm}$ is the initial energy in the center of mass system.  
We choose the central value $0.135\,{\rm GeV}$ of (\ref{eq:Tffit}) 
as $T_f$.  Since the fluid immediately freeze-out at 
$E_{\rm cm}=12\,{\rm GeV}$ as seen from (\ref{eq:gammafit}), 
the value of $\tau_0$ is fixed to reproduce the value 
$dN_{\rm ch}/dy=1.31$ at $E_{\rm cm}=12\,{\rm GeV}$ as seen from 
(\ref{eq:dNdyfit}); 
\begin{eqnarray}
{{dN_{\rm ch}}\over{dy}}
={2\over 3}\tau_0\pi R_0^2\hat\rho_{\pi f}=1.31. \label{eq:tau0fix}
\end{eqnarray}

\begin{figure}
\begin{center}
\includegraphics[width=12truecm]{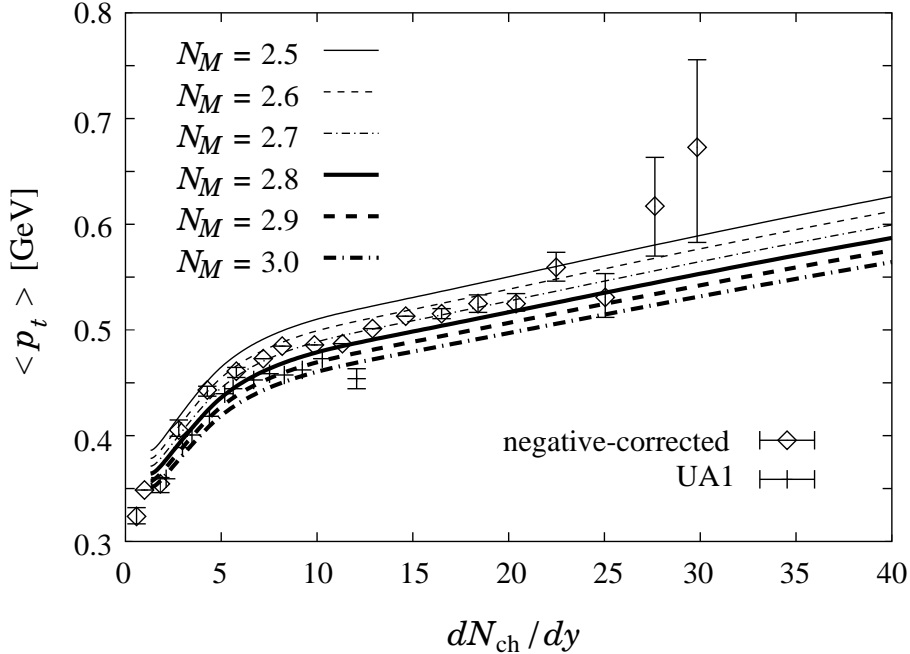}
\caption[]{$\langle p_t\rangle$ is shown as a function of
$dN_{\rm ch}/dy$.  The parameter $N_M$ is searched for the range
$2.5\sim 3.0$ with $B^{1/4}=0.25\,{\rm GeV}$.  The experimental
data are taken from \cite{pt}.}
\end{center}
\end{figure}
Now varying $T_0$ (or $x_0$), we can calculate $\langle p_t\rangle$ 
as a function of $dN_{\rm ch}/dy$ with giving the value of $N_M$.  
The results are shown in Fig.~2.  The value of $N_M$ is searched 
for the range $2.5\sim 3.0$.  The value of 
$\langle p_t\rangle$ is almost determined by 
$\langle m_t^{*2}(\pi_c)\rangle=\langle m_t^{*2}(M)\rangle/N_M$ 
since the value of $N_M\rho_{Mf}$ is almost an order of magnitude 
larger than that of $\rho_{\pi f}$. (See 
(\ref{eq:ptdef})$\sim$(\ref{eq:pic}))  Therefore the calculated 
values of $\langle p_t\rangle$ is sensitive to the value of $N_M$.  
Comparing with the experimental data \cite{pt} we choose as  
\begin{eqnarray}
N_M=2.7,\quad \tau_0=1.29\,{\rm fm}. \label{eq:tau0}
\end{eqnarray} 
\begin{figure}
\begin{center}
\includegraphics[width=12truecm]{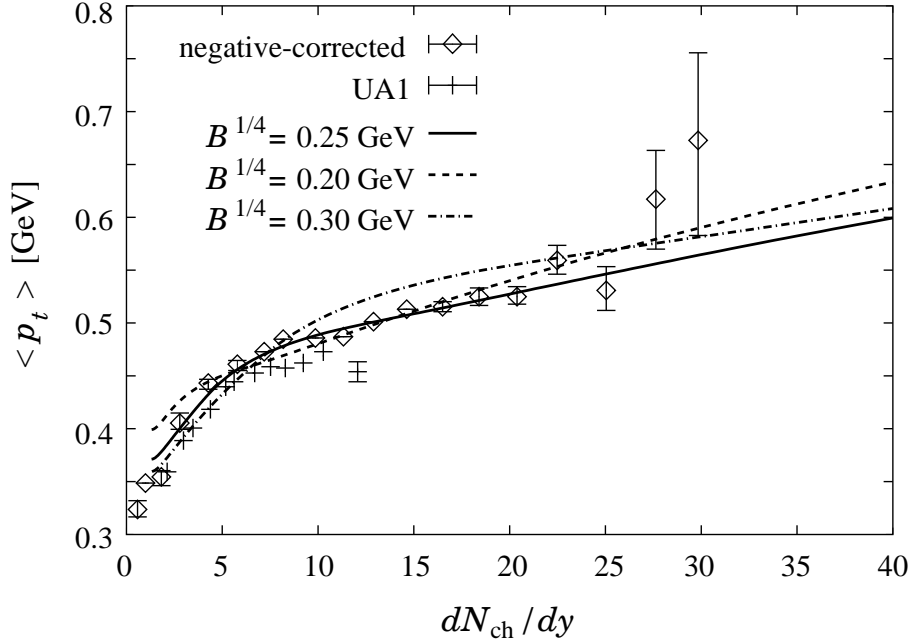}
\caption[]{$\langle p_t\rangle$ is shown as a function of
$dN_{\rm ch}/dy$ with different $B$.  The parameter $N_M$ is set
to $2.7$.  For small $B$, $\langle p_t\rangle$ becomes large
slightly fast as $dN_{\rm ch}/dy$ becomes large.}
\end{center}
\end{figure}
The response to $B$ is shown in Fig.~3.  For small bag constant, 
$\langle p_t\rangle$ becomes large slightly fast as $dN_{\rm ch}/dy$ 
becomes large.  This is recognized as a consequence of a larger 
acceleration for smaller $B$ as is seen from (\ref{eq:dvtq}).  
\begin{figure}
\begin{center}
\includegraphics[width=12truecm]{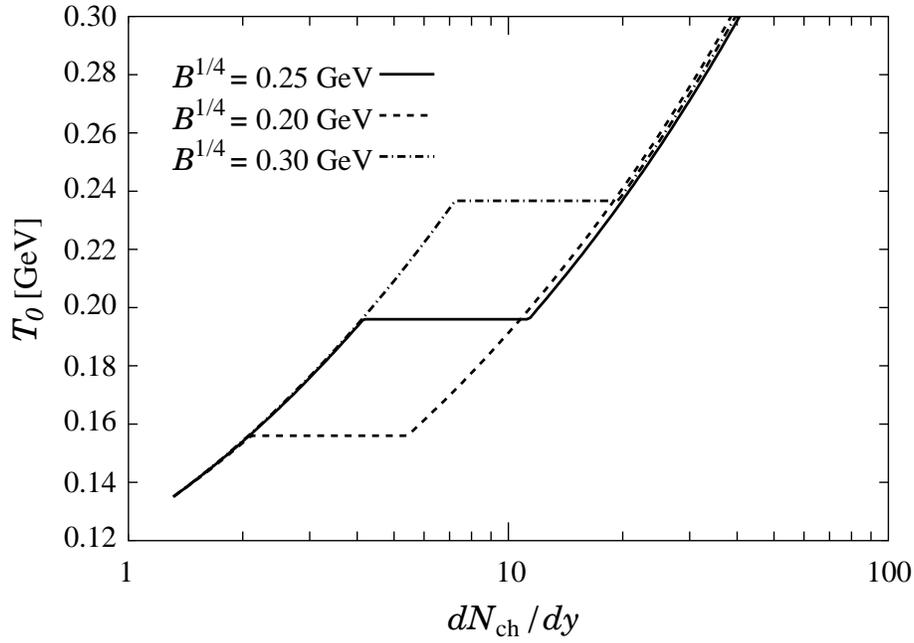}
\caption[]{$T_0$ is shown as a function of $dN_{\rm ch}/dy$.
The parameter $N_M$ is set to $2.7$.
The critical temperature becomes lower and the region of
mixed-phase start becomes narrower for smaller $B$.}
\end{center}
\end{figure}
In Fig.~4, the initial temperature $T_0$ is shown 
as a function of $dN_{\rm ch}/dy$.  For $B^{1/4}=0.25\,{\rm GeV}$ 
the critical temperature is $T_c=0.196\,{\rm GeV}$ and the region 
$4.2<dN_{\rm ch}/dy<11$ corresponds to the mixed-phase start. 
For a larger (smaller) value of $B$, $T_c$ becomes higher (lower) and 
the region of the mixed-phase start becomes wider (narrower).
The value $T_c=0.237\,{\rm GeV}$ for $B^{1/4}=0.30\,{\rm GeV}$ is 
somewhat larger than the value suggested by lattice QCD calculations 
so that we have adopted $B^{1/4}=0.25\,{\rm GeV}$ in (\ref{eq:Bb}).
Comparing with Fig.~3, we can see a slight structure of entrance to mixed 
phase while entrance to quark-gluon phase does not show a remarkable 
structure.  

\begin{figure}
\begin{center}
\includegraphics[width=12truecm]{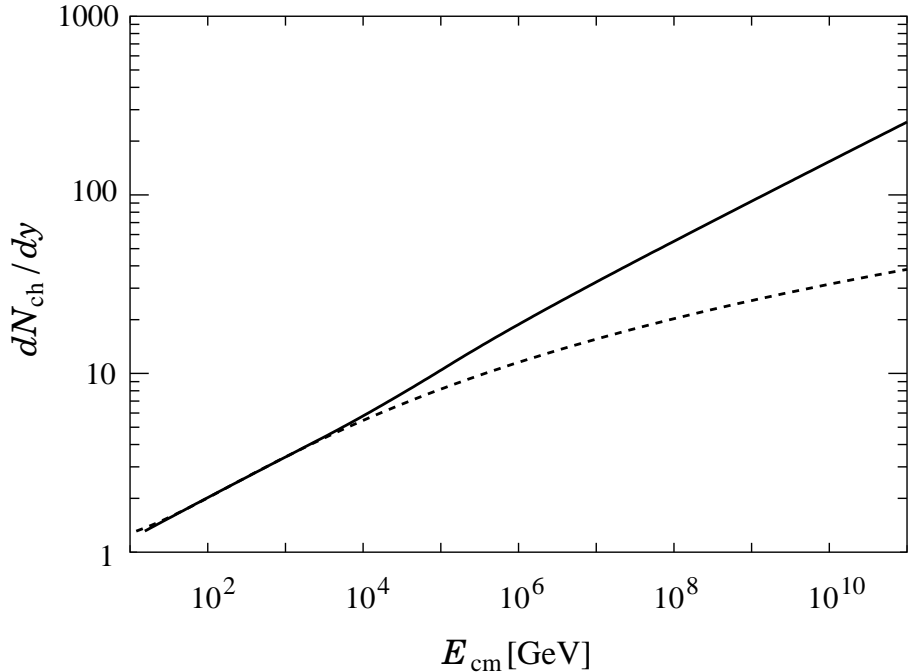}
\caption[]{$dN_{\rm ch}/dy$ is shown as a function of $E_{\rm cm}$.
Solid line show a consequence of (\ref{eq:epsfit}) and (\ref{eq:fitrlt})
while dashed line show (\ref{eq:dNdyfit}).  Both lines are coincide 
below Tevatron region}
\end{center}
\end{figure}
An initial energy density $\varepsilon_0$ has a fundamental importance 
to prescribe an initial state of the fluid and determine its evolution 
in our picture as explained in \S 1.  
The values of $\varepsilon_0$ can be obtained as a function of 
$E_{\rm cm}$ in such a way that the calculated values using 
(\ref{eq:dNchdy}) reproduce the predicted values by (\ref{eq:dNdyfit}).  
However, (\ref{eq:dNdyfit}) is only applicable for 
$E_{\rm cm}\le 1.8\,{\rm TeV}$ since (\ref{eq:dNdyfit}) was obtained 
as a result of fit to experimental data up to Tevatron region.  
In order to extrapolate above Tevatron region we assume the following 
form of $\varepsilon_0$
\begin{eqnarray}
\varepsilon_0(NN)=\varepsilon_{00}
\left({{E_{\rm cm}}\over{E_2}}\right)^\delta, \label{eq:epsfit} 
\end{eqnarray}
and determine $\varepsilon_{00}$ and $\delta$ to reproduce (\ref{eq:dNdyfit}) 
below Tevatron region.  The result is given by 
\begin{eqnarray}
\varepsilon_{00}=0.184\,{\rm GeV/fm^3},\quad \delta=0.294. \label{eq:fitrlt}
\end{eqnarray}
Using this result $dN_{\rm ch}/dy$ is shown as a function of 
$E_{\rm cm}$ in Fig.~5 where solid line show a consequence of 
(\ref{eq:epsfit}) and (\ref{eq:fitrlt}) while dashed line show 
(\ref{eq:dNdyfit}).  We can see both lines are coincide below 
Tevatron region.  

%%%%%%%%%%%
\subsection{$AA$-collision}
\label{AA}
First we set the initial radius $R_0$ for $AA$-collision as 
\begin{eqnarray}
R_0(AA)=1.2 A^{1/3}\,{\rm fm}. \label{R0AA}
\end{eqnarray}
In order to incorporate $AA$ collisions into numerical calculations, 
we need a relation to combine $\varepsilon_0(AA)$ with $\varepsilon_0(NN)$.  
For this purpose we assume that 
\begin{eqnarray}
\varepsilon_0(AA)=A^\beta \varepsilon_0(NN). \label{eq:eps0}
\end{eqnarray}
In the above, both sides should be compared at equal $E_{\rm cm}$ 
which means energy per nucleon in the center of mass system for 
$AA$-collision.  

\begin{figure}
\begin{center}
\includegraphics[width=12truecm]{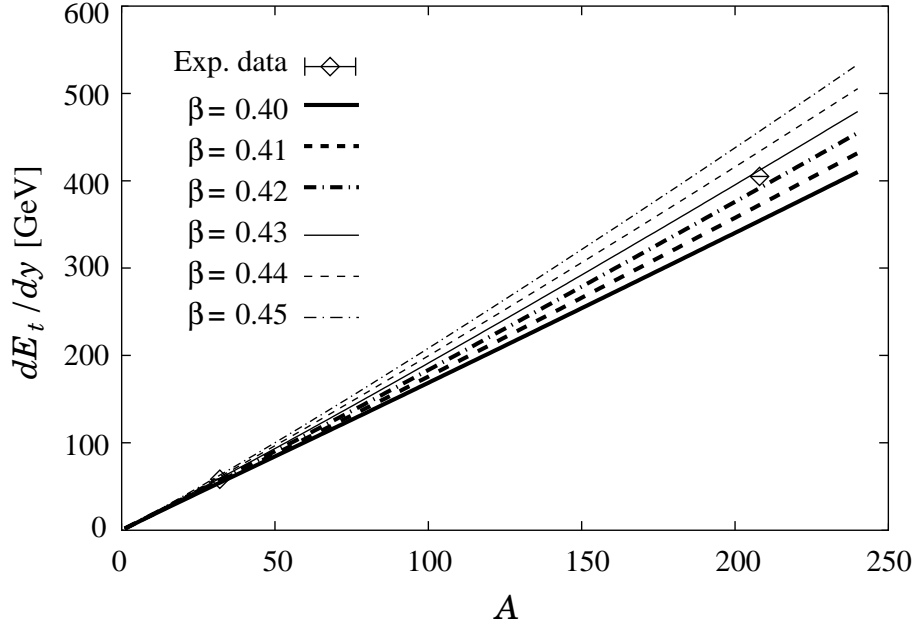}
\caption[]{$dE_t/dy$ is shown as a function of $A$.
The parameter $\beta$ is searched for the range of
$0.40\sim 0.45$.  $E_{\rm cm}$ is set to $20\,{\rm GeV/A}$.
Other parameters are set as
$B^{1/4}=0.25\,{\rm GeV}$, $N_M=2.7$.
Experimental data are taken from \cite{Et}.}
\end{center}
\end{figure}
In Fig.~6 the average transverse energy per unit rapidity interval 
$dE_t/dy$ is shown as a function of $A$ for $E_{\rm cm}=20\,{\rm GeV/A}$ 
and the values of $\beta=0.40\sim 0.45$.  Comparing with experimental 
data \cite{Et}, where the initial energies are 
$E_{\rm cm}=19.4\,{\rm GeV/A}$ ($E_{\rm Lab}=200\,{\rm GeV/A}$) and 
$E_{\rm cm}=17.3\,{\rm GeV/A}$ ($E_{\rm Lab}=158\,{\rm GeV/A}$), 
we choose 
\begin{eqnarray}
\beta=0.43. \label{eq:beta}
\end{eqnarray}
Having fixed all parameters we can now calculate various observables 
as functions of $E_{\rm cm}$ and $A$.  

\begin{figure}
\begin{center}
\includegraphics[width=12truecm]{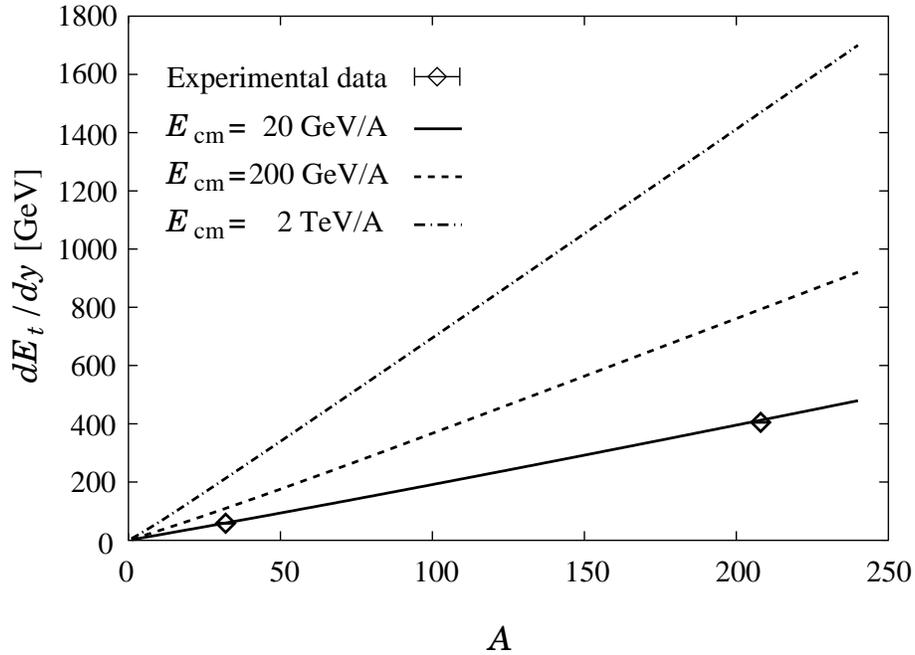}
\caption[]{$dE_t/dy$ is shown as a function of $A$ for various
$E_{\rm cm}$.  The parameters are set as
$B^{1/4}=0.25\,{\rm GeV}$, $N_M=2.7$, $\beta=0.43$.
Experimental data are taken from \cite{Et}.}
\end{center}
\end{figure}
\begin{figure}
\begin{center}
\includegraphics[width=12truecm]{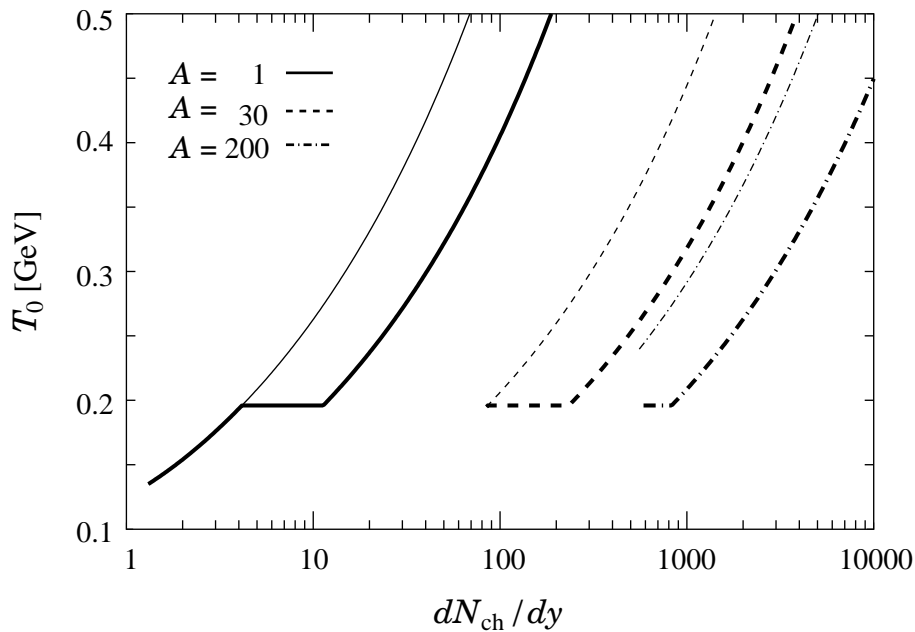}
\caption[]{$T_0$ is shown as a function of $dN_{\rm ch}/dy$.
The parameters are set as
$B^{1/4}=0.25\,{\rm GeV}$, $N_M=2.7$, $\beta=0.43$.
The thin lines show the case of no phase transition
for comparison.}
\end{center}
\end{figure}
In Fig.~7 $dE_t/dy$ is shown as a function of $A$ for 
$E_{\rm cm}=20\,{\rm GeV/A}$, $200\,{\rm GeV/A}$, $2\,{\rm TeV/A}$.   
$dE_t/dy$ grows almost linearly with $A$.  In Fig.~8 the initial 
temperature $T_0$ is shown as a function of $dN_{\rm ch}/dy$ for 
$A=1, 30, 200$.  These results correspond to the experiments for 
various incident energies.  For the case of $A=1$ 
($R_0=R_0(NN)=0.82\,{\rm fm}$) $4.2<dN_{\rm ch}/dy<1.1\times 10$ 
corresponds to mixed-phase start.  
For the case of $A=30$ ($R_0=3.73\,{\rm fm}$)
$8.6\times 10<dN_{\rm ch}/dy<2.3\times 10^2$ corresponds to 
mixed-phase start.  As for the case of $A=200$ ($R_0=7.02\,{\rm fm}$) 
$dN_{\rm ch}/dy<8.2\times 10^2$ corresponds to mixed-phase start.  
The behaviour of $T_0$ in the case of no phase transition is 
also shown by thin lines for comparison.

\begin{figure}
\begin{center}
\includegraphics[width=12truecm]{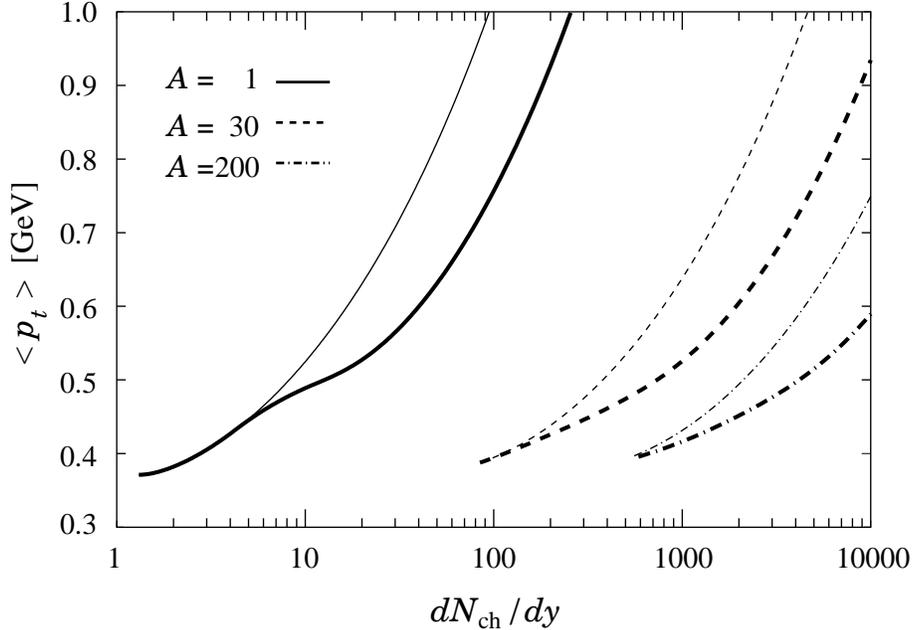}
\caption[]{$\langle p_t\rangle$ is shown as a function of
$dN_{\rm ch}/dy$.  The parameters are set as
$B^{1/4}=0.25\,{\rm GeV}$, $N_M=2.7$, $\beta=0.43$.
The thin lines show the case of no phase transition for comparison.}
\end{center}
\end{figure}
In Fig.~9 the average transverse momentum $\langle p_t\rangle$ is 
shown as a function of $dN_{\rm ch}/dy$ for $A=1, 30, 200$.  
These results also correspond to the experiments for various 
incident energies.  The behaviour of $\langle p_t\rangle$ in the 
case of no phase transition is also shown by thin lines for comparison.  
It is clearly seen that the fluid is not accelerated so much in the
mixed phase.  The reason is that the acceleration in the mixed phase
given by the right hand side of (\ref{eq:dvtm}) has no $1/\tau$-term 
and an extra factor $(1-v_t^2)$ in $1/R$-term so that it takes 
smaller values than the acceleration in hadron/quark-gluon phase 
given by (\ref{eq:dvt})/(\ref{eq:dvtq}).  This is a common feature 
to the case of massless hadron, which is supposed to be pion, discussed 
in the previous paper \cite{fluid}.  The difference lies in the fact 
that the interval of mixed-phase start in the present case is narrower than 
the one in the case of a single massless hadron so that the structure 
is less prominent especially for $A=30, 200$.  This is because the 
degree of freedom in the hadron phase is larger for the present case 
which takes into account the contribution from continuum of meson 
resonances.  However it should be noticed that the entrance to 
quark-gluon phase can be seen in this figure plotted in the log 
scale of $dN_{\rm ch}/dy$ in contrast to Fig.~3 plotted in the 
linear scale.  

\begin{figure}
\begin{center}
\includegraphics[width=12truecm]{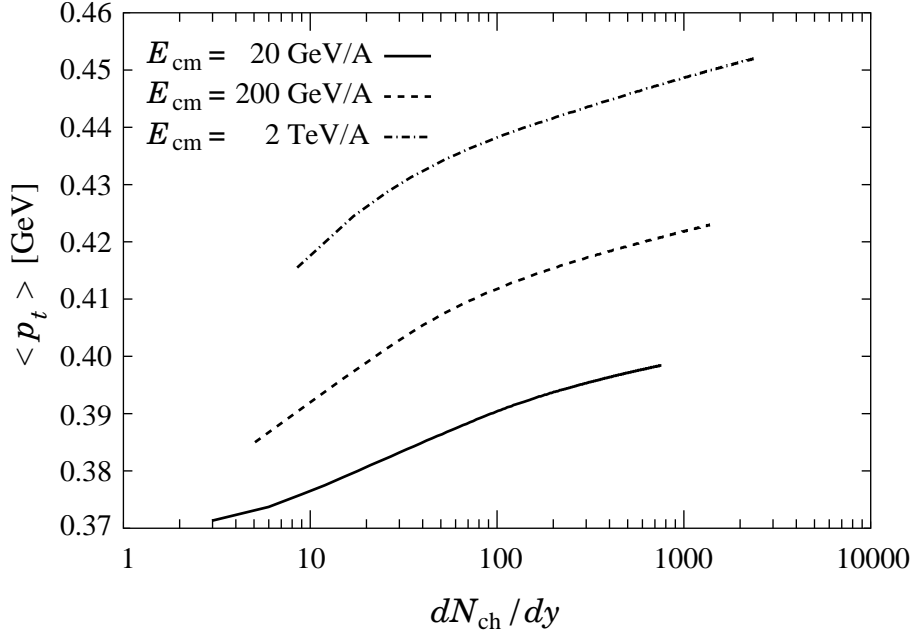}
\caption[]{$\langle p_t\rangle$ is shown as a function of 
$dN_{\rm ch}/dy$ with varying $A=1\sim 240$.
The parameters are set as
$B^{1/4}=0.25\,{\rm GeV}$, $N_M=2.7$, $\beta=0.43$.}
\end{center}
\end{figure}
\begin{figure}
\begin{center}
\includegraphics[width=12truecm]{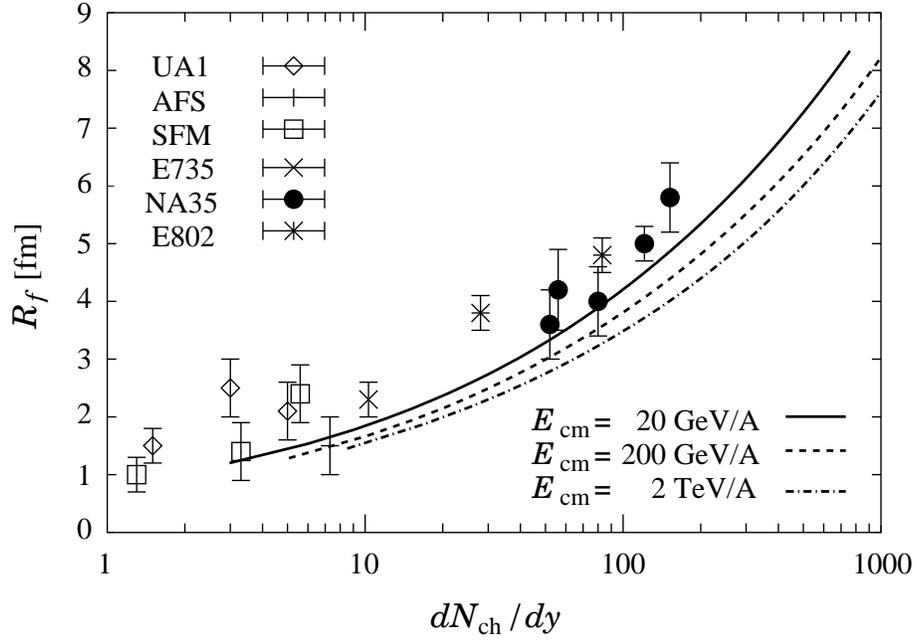}
\caption[]{$R_f$ is shown as a function of $dN_{\rm ch}/dy$.
The parameters are set as
$B^{1/4}=0.25\,{\rm GeV}$, $N_M=2.7$, $\beta=0.43$.
The experimental data are taken from \cite{Rf}.}
\end{center}
\end{figure}
In Fig.~10 $\langle p_t\rangle$ is shown as a function of $dN_{\rm ch}/dy$ 
for $E_{\rm cm}=20\,{\rm GeV}, 200\,{\rm GeV}, 2\,{\rm TeV}$ with varying 
$A$ from 1 to 240.  To this variation the response of 
$\langle p_t\rangle$ is mild and does not become so large even 
for $E_{\rm cm}=2\,{\rm TeV}$.  
In Fig.~11 the final radius $R_f$ is shown as a function of 
$dN_{\rm ch}/dy$ for $E_{\rm cm}=20\,{\rm GeV}, 200\,{\rm GeV}, 2\,{\rm TeV}$ 
with varying $A$ from 1 to 240.  The experimental data are taken 
from \cite{Rf}.  The calculated values does not differ so much from 
the observed values.  It should be noticed that $R_f$ is rather 
insensitive to the variation of $E_{\rm cm}$.  As for the response to 
the variation of $\beta$, $R_f$ becomes larger for smaller $\beta$.  

\begin{figure}
\begin{center}
\includegraphics[width=12truecm]{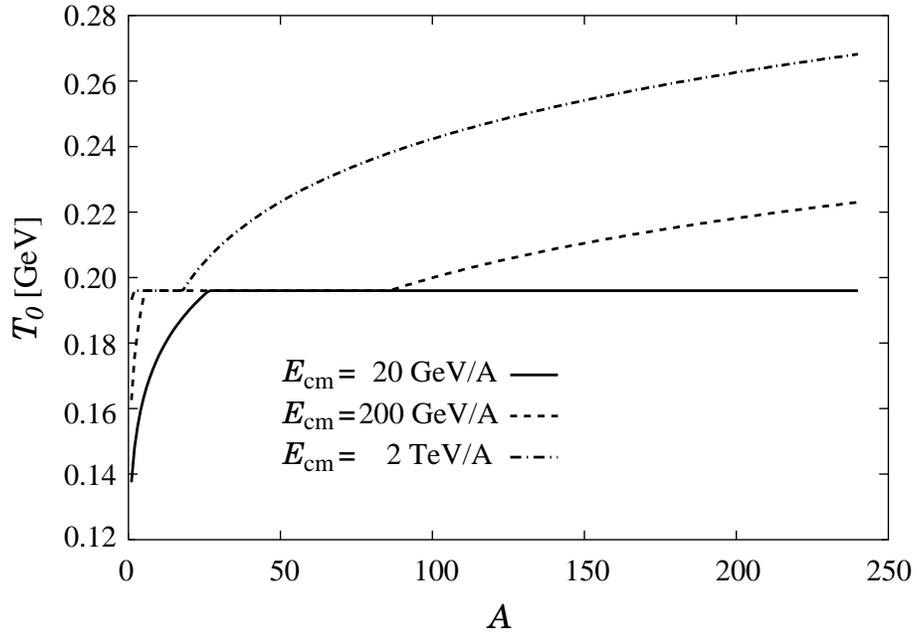}
\caption[]{$T_0$ is shown as a function of $A$.
The parameters are set as
$B^{1/4}=0.25\,{\rm GeV}$, $N_M=2.7$, $\beta=0.43$.}
\end{center}
\end{figure}
Finally in Fig.~12 the initial temperature is shown as a function of 
$A$ for $E_{\rm cm}=20\,{\rm GeV},$\break$200\,{\rm GeV}, 2\,{\rm TeV}$.  
For $E_{\rm cm}=20\,{\rm GeV}$ the entrance to the mixed phase occurs 
at $A=27$ and the entrance to the quark-gluon phase does not occur even 
at highest $A$.  For $E_{\rm cm}=200\,{\rm GeV}$ the entrance to the 
mixed phase occurs at $A=6$ and the entrance to the quark-gluon phase 
occurs at $A=87$.  As for $E_{\rm cm}=2\,{\rm TeV}$ the entrance to the 
mixed phase occurs at $A=2$ and the entrance to the quark-gluon phase 
occurs at $A=18$.  Thus quark-gluon phase will be realized at BNL RHIC 
energies while it is not at CERN SPS energies.  

%%%%%%%%%%%%%%%%%
\section{Conclusions and discussion}
\label{sammary}
We have shown that our simplified fluid model easily calculates 
observables in nucleon-nucleon and nucleus-nucleus collisions at 
high energy region.  As for the equation of state for hadron phase, 
we have adopted the compressible bag model.  This model can satisfy 
both requirements that one should take into account an infinite degrees 
of freedom of hadrons and that deconfinement transition should be take 
place, which usually contradict mutually in models other than 
compressible bag model.  For an evolution 
of the fluid, we have assumed the cylindrical symmetry with respect 
to the collision axis and solved the averaged quantities over the 
transverse plane.

In the previous paper \cite{fluid} of the averaged evolution equation, 
we used the equation of state of massless pion for hadron phase.  Both 
calculations give the result that agrees well in terms of 
qualitative analysis.  However, the present approach produces 
$(\rho_{\pi f}+N_M\rho_{Mf})/\rho_{\pi f}$ times of pions in
comparison with the previous one.  This gives a reasonable 
value for $\tau_0$  by the condition of (\ref{eq:tau0fix}).  
On the other hand, the ratio $\varepsilon_Q(T_c)/\varepsilon_H(T_c)$ 
becomes smaller because of the increase of hadronic degrees of freedom.

The increase of the average transverse momentum is brought about 
by a transverse flow of the fluid.  As is seen from (\ref{eq:dvt}) 
the pressure slope $p/R$ in the transverse direction mainly causes 
the transverse flow.  In the mixed phase the effect from the 
corresponding term became smaller as shown by equation (\ref{eq:dvtq}).  
This gives a structure of an average transverse momentum as a 
function of $dN_{\rm ch}/dy$. In the hadron and the quark-gluon 
phase the ratio of the pressure to the energy density increases 
rapidly with the increase of the energy density.  The ratio in 
both phases reaches $1/3$ in high energy density limit, so that 
the acceleration is generated similarly in both phases as shown 
in Fig.~9.  In the case of the quark-gluon phase start the 
acceleration becomes weaker because of passing through the 
mixed phase.  As for $AA$-collisions the acceleration becomes 
weaker with an increase of $A$ because of a decrease of the 
pressure slope in the transverse direction.  

As shown in Figs.~8 and 9 $dN_{\rm ch}/dy$ is not a reasonable 
variable for plotting the observables in $AA$-collisions except for 
the transverse source radius. In our model the equations of 
(\ref{eq:dvt}) and (\ref{eq:eps0}) generate the $A$ dependence.  
We have to find a scaling variable in $AA$-collision.

We have assumed the relation between the initial energy densities 
in $NN$- and $AA$- collisions by equation (\ref{eq:eps0}).  
This power-law assumption is similar to the relations given 
by the dilute-gas and multiple-collision models \cite{fluid}.  
Finally we predict that the quark-gluon plasma can be observed 
for average multiplicity events in $AA$-collision with $A\ge 87$ 
at RHIC energies.  The numerical calculation results are shown 
in Fig.~12.

%%%%%%%%%%%%%%%%

\end{document}